# Exploring Graphene Effect on Fiber Optic Surface Plasmon Resonance Biosensors

**Mariam M. Moussilli[1], Abdul Rahman El Falou[1], Raed M. Shubair[2]**

[1] Department of Electrical and Computer Engineering, Beirut Arab University (BAU), Lebanon

[2] Research Laboratory Of Electronics, Massachusetts Institute of Technology (MIT), USA

**Abstract**—This report investigates design enhancements of Fiber Optic Surface Plasmon Resonance (FO-SPR) biosensors with Gold and Silver metallic films. The effect of adding a Graphene coating on the sensitivity and detection accuracy of FO-SPR biosensors is studied. We also compare the results for the addition of five layers of Graphene on both Gold-based and Silver-based FO-SPR biosensors in order to determine the optimum sensor design. Our results indicate that the sensitivity of the biosensor increases when a Graphene coating is added, while the detection accuracy decreases. Results also demonstrate that the sensitivity due to adding Graphene sheets increases more for Silver-based biosensor when compared to Gold-based biosensors.

## [1]    INTRODUCTION

This report is concerned with Fiber Optic Surface Plasmon Resonance (FO-SPR) biosensors which are label-free optical biosensors used to measure the concentration of a desired analyte in a sample. These biosensors have surged in popularity in recent years due to their high sensitivity, nano-size, light weight, simplistic design, adaptation to harsh environments, possibility of remote sensing, impossibility of electrocution, immunity to electromagnetic interference, resistance to corrosion, and need for minute sample quantities and conservation of the sample integrity and minimal invasiveness which are essential requirements in medical applications [1] [2] [3].  The work in this project is a continuation of the several previous contributions which have been reported in the literature in various relevant technologies, systems, and their associated applications [9-17] and [18-27].  The report is organized as follows: Section II explains the operating principle of FO-SPR sensors and highlights the sensitivity and and detection accuracy as the key performance parameters of FO-SPR sensors.  Section III includes the simulation results, which demonstrate the effect of adding Graphene coating on a Gold-based and a Silver-based FO-SPR sensors.  Finally, conclusions are outlined in Section IV.

**[2]      FO-SPR OPERATION AND PERFORMANCE PARAMETERS**

In FO-SPR sensors, Surface Plasmon Waves (SPWs) are excited by absorbed evanescent waves. SPWs transduce the change in the Surrounding Refractive Index (SRI) of the sample into a change in the optical signal received at the Optical Spectrum Analyzer (OSA). Upon SRI variation, the change in the signal received at the OSA is represented as a shift in the wavelength where maximum power is absorbed known as the resonance wavelength ($\lambda_{res}$) [4] [5] [6]. In Graphene enhanced FO-SPR biosensors, a Graphene coating is added to the metallic layer of the sensor probe in order to improve the performance of the sensor [7] [8]. The $S_n$ is the measure of the ability of the sensor to translate the variation in the SRI of the sample into a shift in the $\lambda_{res}$ at the OSA [8] [9]. The DA is the measure of the ability of the sensor to pinpoint the exact $\lambda_{res}$ by calculating the depth of the dip in the power at the $\lambda_{res}$ with respect to the width of the dip [1] [8]. The work in this project is a continuation of the several previous contributions, which have been reported in the literature in various relevant technologies, systems, and their associated applications [9-27].

## [3]     ADDITION OF GRAPHENE TO THE GOLD AND SILVER SENSORS

We follow the 4-layer model presented in [2] [7] [8] in order to simulate the response curves of the FO-SPR sensor for varying SRI= [1.33,1.35,1.37]. The values used are as follows: the core diameter is 100µm, the core refractive index is 1.451, the cladding refractive index is 1.450, the sensing region length is 10mm, the metal thickness is 40nm and with 5 Graphene sheets.

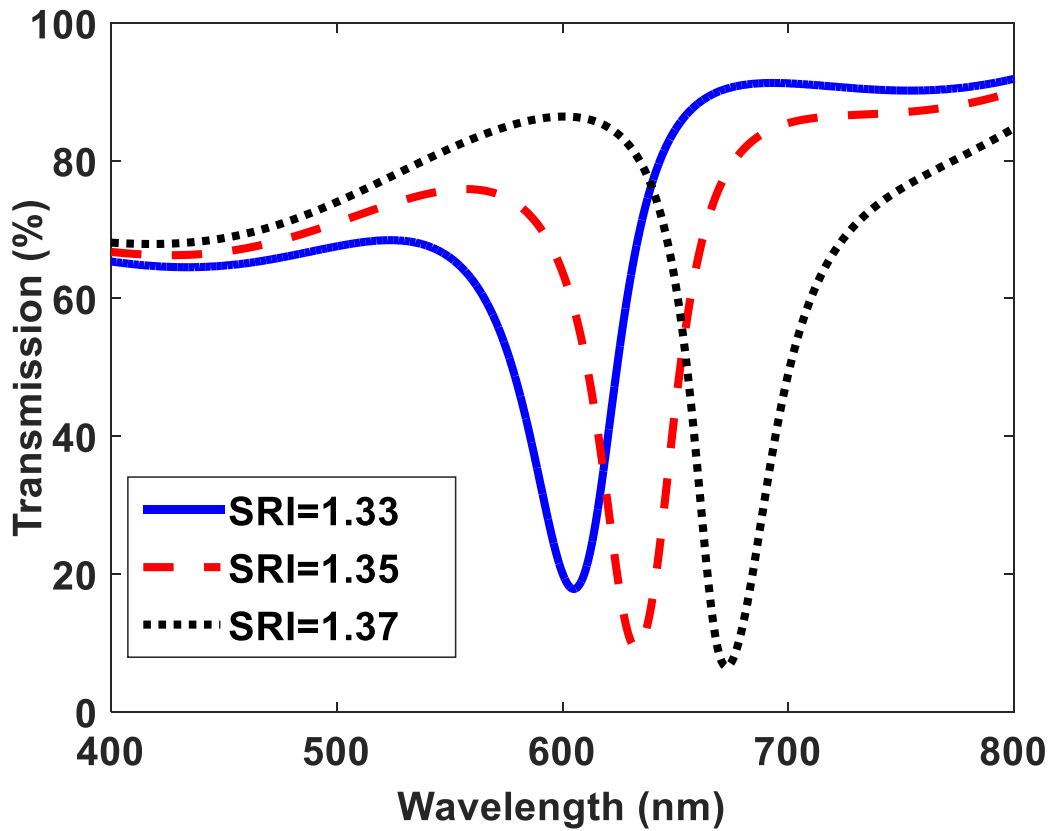

Figure 1-Gold-based SPR sensor without Graphene for varying SRI

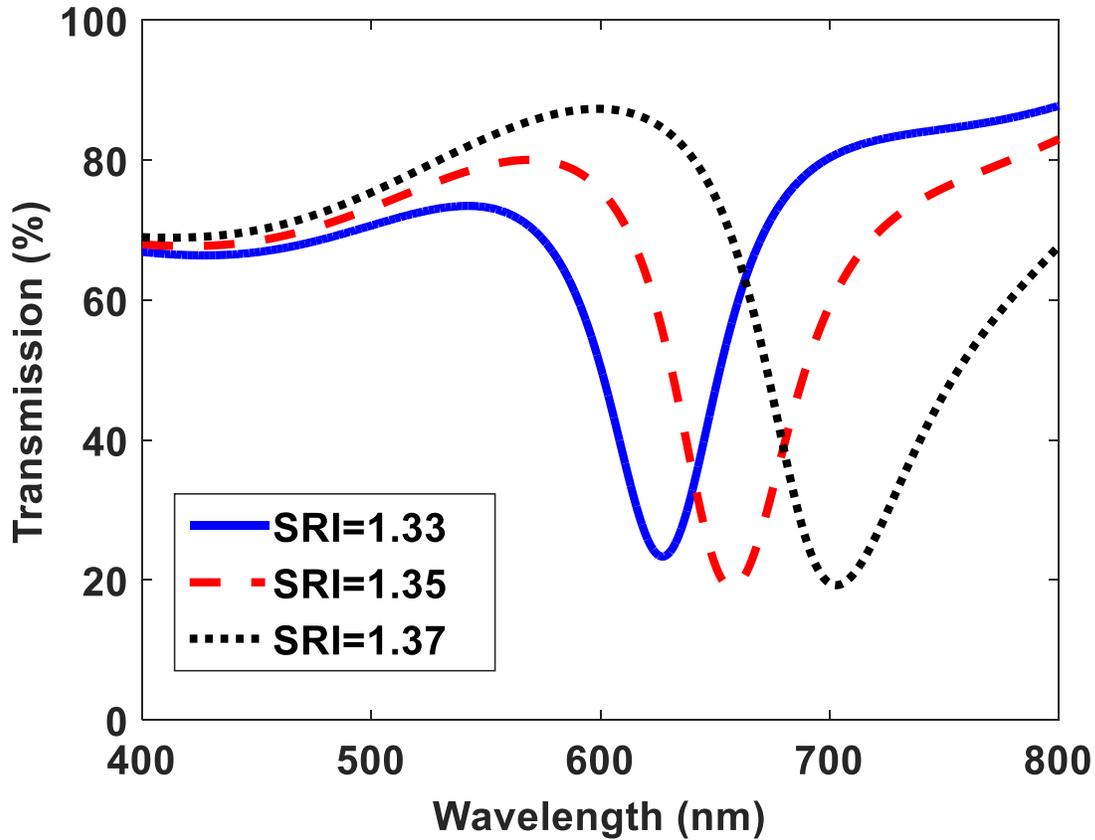

Figure 2-Gold-based SPR sensor with Graphene for varying SRI

When comparing fig.1 and fig.2 for increasing SRI, the red shift in the $\lambda_{res}$ becomes greater and the response curves become shallower and broader when Graphene is added to the Gold sensor. So we can conclude that the sensitivity increases and the detection accuracy decreases upon the addition of Graphene which is also assured by the calculations in table 1. From table 1 and at SRI=1.35, the addition of Graphene increases the Sn by 8.98% and decreases the DA by 36.82%.

| SRI | 1.33 | 1.35 | 1.37 |
|---|---|---|---|
| **Before the Addition of Graphene** | | | |
| $S_n$ (Nm/RIU) | Calibration Curve | 1392.5 | 1691.25 |
| DA (TU%/nm) | 1.7673 | 2.1973 | 2.0470 |
| **After Addition of 5 Graphene Layers** | | | |
| $S_n$ (Nm/RIU) | Calibration Curve | 1517.5 | 1895 |
| DA (TU%/nm) | 1.4811 | 1.3882 | 0.9669 |

Table 1-Performance parameters for the Gold SPR sensor

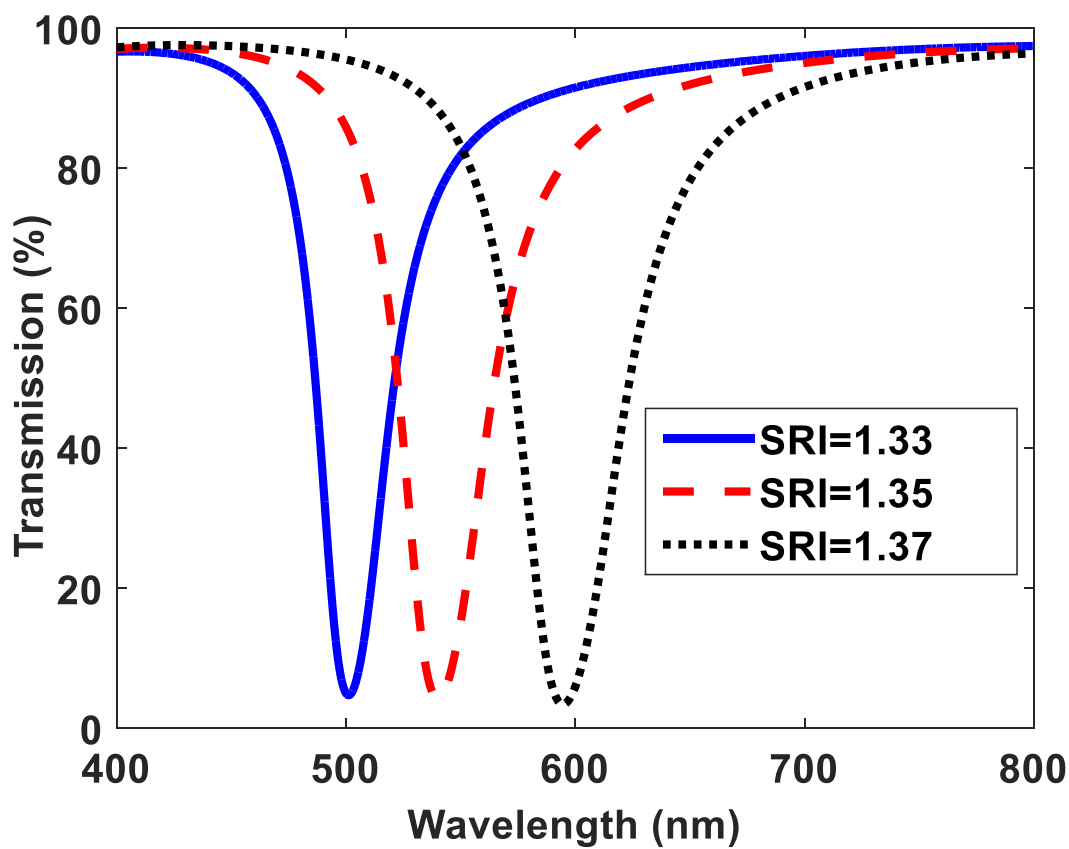

Figure 3-Silver-based SPR sensor without Graphene for varying SRI

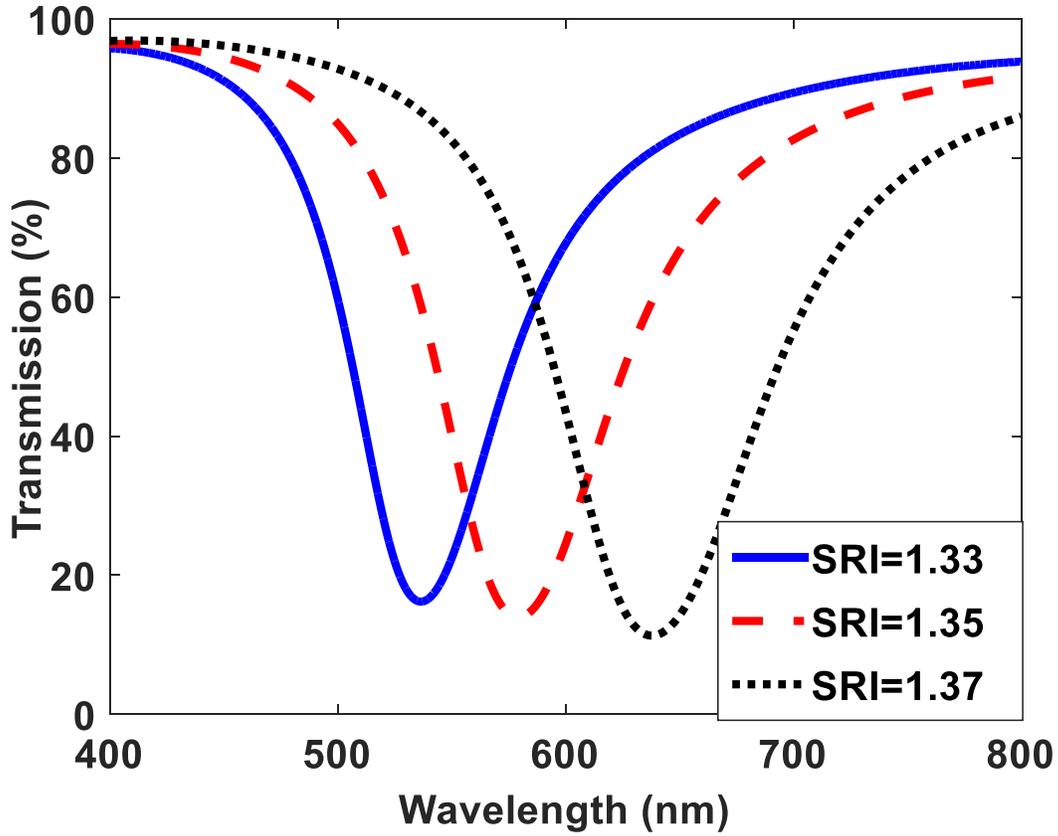

ct

Figure 4-Silver-based SPR sensor with Graphene for varying SRI

| SRI | 1.33 | 1.35 | 1.37 |
|---|---|---|---|
| **Before the Addition of Graphene** | | | |
| $S_n$ (nm/RIU) | Calibration Curve | 1947.5 | 2337.5 |
| DA (TU%/nm) | 2.7373 | 2.2386 | 1.9122 |
| **After Addition of 5 Graphene Layers** | | | |
| $S_n$ (nm/RIU) | Calibration Curve | 2130 | 2535 |
| DA (TU%/nm) | 1.2033 | 1.0400 | 0.9010 |

Table 2-Performance parameters for the Silver SPR sensor

Based on fig.3 and fig.4, the same observations for the Gold sensor stand for the Silver sensor and these observations are assured by the calculations presented in table 2. From table 2 and at SRI=1.35, the addition of Graphene increases the Sn by 9.37% and decreases the DA by 53.54%.

Regarding sensors without a Graphene layer, we deduce that a Silver sensor exhibits higher $S_n$ and higher DA compared to a Gold sensor of same parameters. For sensors with a Graphene layer, we deduce that the $S_n$ increase of the Silver sensor is greater than the $S_n$ increase of the Gold sensor and that the DA decrease of the Silver sensor is greater than the DA decrease of the Gold sensor noting that despite this, the DA of the Graphene enhanced silver sensor remains higher than the DA of the Graphene enhanced Gold sensor.

## [4] CONCLUSION

Our investigations have shown that adding Graphene coating to a FO-SPR biosensor increases sensor sensitivity significantly but with the trade-off of decreasing the sensor's detection accuracy. These parameters predominantly improve for FO-SPR sensors based on Silver when compared to Gold. The work in this project is a continuation of the several previous contributions, which have been reported in the literature in various relevant technologies, systems, and their associated applications [9]-[27].

**REFERENCES**

[1]     C. Fallauto, G. Perrone and A. Vallan, "Impact of Optical Fiber Characteristics in SPR Sensors for Continuous Glucose Monitoring," in IEEE MeMeA2014, Libson, 2014.

[2]     H. Fu, S. Zhang, H. Chen and J. Weng, "Graphene Enhances the Sensitivity of Fiber-Optic Surface Plasmon Resonance Biosensor," IEEE SENSORS JOURNAL, VOL. 15, NO. 10, pp. 5478-5482, 2015.

[3]     J. Homola, S. S. Yee and G. Gauglitz, "Surface plasmon resonance sensors: review," Sensors and Actuators B 54, pp. 3-15, 1999.

[4]     M. M. Moussilli, A. R. El Falou and R. M. Shubair, "Overview of Fiber Optic Surface Plasmon Resonance Biosensors for Medical Applications," in IEEE ANTEM2018, Waterloo, 2018.

[5]     A. K. Sharma, R. Jha and B. D. Gupta, "Fiber-Optic Sensors Based on Surface Plasmon," IEEE Sensors Journal, vol. 7, no. 8, pp. 1118-1129, August 2007.

[6]     B. Gupta and R. Verma, "Surface Plasmon Resonance-Based Fiber Optic Sensors: Principle, Probe Designs, and Some Applications," Journal of Sensors, pp. 1-4, 2009.

[7]     Y. Saad, M. Selmi, M. H. Gazzah and H. Belmabrouk, "Graphene Effect on the Improvement of the Response of Optical Fiber SPR Sensor," IEEE SENSORS JOURNAL, VOL. 17, NO. 22, 15 November 2017.

[8]     M. M. Moussilli, A. R. El Falou and R. M. Shubair, "On the Enhancement of Surface Plasmon Resonance Biosensors by Graphene for Medical Applications," in IEEE ANTEM2018, Waterloo, 2018.

[9]     P. Arasu, A. Shabaneh, S. Girei, M. Yaacob and A. Noor, "Enhancement of fiber-SPR sensor utilizing Graphene oxide," IEEE Sensors Journal, pp. 148-149, 2013.

[10]    A. Omar and R. Shubair, "UWB coplanar waveguide-fed-coplanar strips spiral antenna," in 2016 10th European Conference on Antennas and Propagation (EuCAP), Apr. 2016, pp. 1–2.

[11]    H. Elayan, R. M. Shubair, J. M. Jornet, and P. Johari, "Terahertz channel model and link budget analysis for intrabody nanoscale communication," IEEE transactions on nanobioscience, vol. 16, no. 6, pp. 491–503, 2017.

[12]    H. Elayan, R. M. Shubair, and A. Kiourti, "Wireless sensors for medical applications: Current status and future challenges," in Antennas and Propagation (EUCAP), 2017 11th European Conference on. IEEE, 2017, pp. 2478–2482.

[13]    H. Elayan and R. M. Shubair, "On channel characterization in human body communication for medical monitoring systems," in Antenna Technology and

Applied Electromagnetics (ANTEM), 2016 17th International Symposium on. IEEE, 2016, pp. 1–2.

[14]   H. Elayan, R. M. Shubair, A. Alomainy, and K. Yang, "In-vivo terahertz em channel characterization for nano-communications in wbans," in Antennas and Propagation (APSURSI), 2016 IEEE International Symposium on. IEEE, 2016, pp. 979–980. 35

[15]   H. Elayan, R. M. Shubair, and J. M. Jornet, "Bio-electromagnetic thz propagation modeling for in-vivo wireless nanosensor networks," in Antennas and Propagation (EUCAP), 2017 11th European Conference on. IEEE, 2017, pp. 426–430.

[16]   H. Elayan, C. Stefanini, R. M. Shubair, and J. M. Jornet, "End-to-end noise model for intra-body terahertz nanoscale communication," IEEE Transactions on NanoBioscience, 2018.

[17]   H. Elayan, P. Johari, R. M. Shubair, and J. M. Jornet, "Photothermal modeling and analysis of intrabody terahertz nanoscale communication," IEEE transactions on nanobioscience, vol. 16, no. 8, pp. 755–763, 2017.

[18]   H. Elayan, R. M. Shubair, J. M. Jornet, and R. Mittra, "Multi-layer intrabody terahertz wave propagation model for nanobiosensing applications," Nano Communication Networks, vol. 14, pp. 9–15, 2017.

[19]   H. Elayan, R. M. Shubair, and N. Almoosa, "In vivo communication in wireless body area networks," in Information Innovation Technology in Smart Cities. Springer, 2018, pp. 273–287.

[20]   M. O. AlNabooda, R. M. Shubair, N. R. Rishani, and G. Aldabbagh, "Terahertz spectroscopy and imaging for the detection and identification of illicit drugs," in Sensors Networks Smart and Emerging Technologies (SENSET), 2017, 2017, pp. 1–4. 36

[21]   M. S. Khan, A.-D. Capobianco, A. Iftikhar, R. M. Shubair, D. E. Anagnostou, and B. D. Braaten, "Ultra-compact dual-polarised UWB MIMO antenna with meandered feeding lines," IET Microwaves, Antennas & Propagation, vol. 11, no. 7, pp. 997–1002, Feb. 2017.

[22]   M. S. Khan, A. Capobianco, S. M. Asif, D. E. Anagnostou, R. M. Shubair, and B. D. Braaten, "A Compact CSRR-Enabled UWB Diversity Antenna," IEEE Antennas and Wireless Propagation Letters, vol. 16, pp. 808–812, 2017.

[23]   R. M. Shubair, A. M. AlShamsi, K. Khalaf, and A. Kiourti, "Novel miniature wearable microstrip antennas for ISM-band biomedical telemetry," in 2015 Loughborough Antennas Propagation Conference (LAPC), 2015, pp. 1–4.

[24]    R. M. Shubair and H. Elayan, "In vivo wireless body communications: State-of-the-art and future directions," in Antennas & Propagation Conference (LAPC), 2015 Loughborough. IEEE, 2015, pp. 1–5.

[25]    R. M. Shubair and H. Elayan, "Enhanced WSN localization of moving nodes using a robust hybrid TDOA-PF approach," in 2015 11th International Conference on Innovations in Information Technology (IIT), Nov. 2015, pp. 122–127.

[26]    H. Elayan and R. M. Shubair, "Towards an Intelligent Deployment of Wireless Sensor Networks," in Information Innovation Technology in Smart Cities. Springer, Singapore, 2018, pp. 235–250.

[27]    H. Elayan and R. M. Shubair, "Improved DV-hop localization using node repositioning and clustering," in 2015 International Conference on Communications, Signal Processing, and their Applications (ICCSPA), Feb. 2015, pp. 1–6.